\documentstyle[12pt,amsfonts]{article}
\def\one{1\hskip-.37em 1}

\def\half{\textstyle{\frac{1}{2}}}

\def\H{{\cal H}}
\def\D{{\cal D}}
\def\s{\hskip.013cm}

\def\E{{\rm I}\hskip-.2em{\rm E}}
\def\ra{\rightarrow}
\def\tint{{\textstyle\int}}
\def\d{\partial}
\def\o{\overline}

\def\b{\begin{eqnarray*}}     
\def\e{\end{eqnarray*}}       
\def\bn{\begin{eqnarray}}     
\def\en{\end{eqnarray}}       
\def\<{\langle}
\def\>{\rangle}

\def\no{\nonumber}

\def\{{\lbrace}
\def\}{\rbrace}
\def\hg{{\hat g}}
\def\hp{{\hat\pi}}
\def\d{\delta}
\begin{document}
\date{}
\title{The Affine Quantum Gravity Program}
\author{John R. Klauder\footnote{Electronic mail: klauder@phys.ufl.edu}\\
Departments of Physics and Mathematics\\
University of Florida\\
Gainesville, FL  32611}

\maketitle

\begin{abstract}

The central principle of affine quantum gravity is securing and maintaining 
the strict positivity of the matrix $\{\hg_{ab}(x)\}$ composed of the 
spatial components of the local metric operator. On spectral grounds, 
canonical commutation relations are incompatible with this principle, 
and they must be replaced by noncanonical, affine commutation relations. 
Due to the partial second-class nature of the quantum gravitational 
constraints, it is advantageous to use the recently developed projection 
operator method, which treats all quantum constraints on an equal footing. 
Using this method, enforcement of regularized versions of the gravitational 
operator constraints is formulated quite naturally by means of a novel and 
relatively well-defined functional integral involving only the same set of 
variables that appears in the usual classical formulation. It is 
anticipated that skills and insight to study this formulation can be 
developed by studying special, reduced-variable models that still retain 
some basic characteristics of gravity, specifically a partial second-class 
constraint operator structure. Although perturbatively nonrenormalizable, 
gravity may possibly be understood nonperturbatively from a hard-core 
perspective that has proved valuable for specialized models. Finally, 
developing a procedure to pass to the genuine physical Hilbert space 
involves several interconnected steps that require careful coordination. 
\end{abstract}

\subsection*{INTRODUCTION}
Despite being a very difficult problem, quantization of the gravitational 
field has attracted considerable attention because of its fundamental
importance. Among the currently favored approaches to quantize gravity
 is work associated with string theory, or work that is part of 
the canonical program, both important schemes; see, e.g., \cite{str,ash}. 

A relatively new effort to study quantum gravity is the affine quantum 
gravity program the highlights of which are reviewed in this article and 
which is also intended to introduce this program to those who are not 
familiar with it. Although the principles involved are quite conservative 
and fairly natural, this program nevertheless involves a somewhat 
unconventional approach when compared with more traditional techniques. 
(Several precursors to the present program are briefly discussed in 
\cite{kla1}, while details are available in \cite{kla1,kla2, kla3}.)

\subsubsection*{Basic principles of affine quantum gravity}
 
The program of affine quantum gravity is founded on {\it four basic
principles} which we briefly review here. {\it First}, like the 
corresponding classical variables, the 6 components of the spatial 
metric field operators $\hg_{ab}(x)\,[=\hg_{ba}(x)]$, $a,b=1,2,3$,  
form a {\it positive-definite $3\!\times\! 3$ matrix for all $x$}. 
{\it Second}, to ensure self-adjoint kinematical variables when smeared, 
it is necessary to adopt the {\it affine commutation relations} (with 
$\hbar=1$)
\bn &&[\hp^a_b(x),\hp^c_d(y)]=i\half[\d^c_b\,\hp^a_d(x)-\d^a_d\,
\hp^c_b(x)]\,\d(x,y)\;,\no\\ 
  && \hskip-.08cm[\hg_{ab}(x),\hp^c_d(y)]=i\half[\d^c_a\,\hg_{db}(x)
+\d^c_b\,\hg_{ad}(x)]\,\d(x,y)\;,\\
  && \hskip-.18cm[\hg_{ab}(x),\hg_{cd}(y)]=0\;\no  \label{e1}\en
between the metric and the 9 components of the ``scale'' field operator 
$\hp^a_b(x)$. {\it Third}, the principle of {\it quantization before any 
constraints are introduced}, due to Dirac, strongly suggests that the 
basic fields $\hg_{ab}$ and $\hp^c_d$ are initially realized by 
{\it ultralocal representations}, which is explained below. {\it Fourth}, 
and last, introduction and enforcement of the gravitational constraints 
not only leads to the physical Hilbert space but it has the added virtue 
that all vestiges of the temporary ultralocal representation vanish and 
are replaced by physically acceptable alternatives. 
In attacking these basic issues full use of {\it coherent state methods} 
and the {\it projection operator method} for constrained system 
quantization is made.

The affine coherent states are defined (for $\hbar=1$) by
  \bn |\pi,\gamma\>\equiv e^{i\tint \pi^{ab}\hg_{ab}\,d^3\!x}\,
e^{-i\tint\gamma^a_b\hp^b_a\,d^3\!x}\,|\eta\> \label{e2}\en
for general, smooth, $c$-number fields $\pi^{ab}\,[=\pi^{ba}]$ and 
$\gamma^c_d$ of compact support, and are chosen so that the coherent 
state overlap function becomes 
 \bn  &&\hskip-.5cm\<\pi'',\gamma''|\pi',\gamma'\>= \exp
\bigg(\!-\!2\int b(x)\,d^3\!x\, \no\\
  &&\hskip.5cm\times\ln\bigg\{  \frac{
\det\{\half[g''^{kl}(x) +g'^{kl}(x)]+i\half b(x)^{-1}[\pi''^{kl}(x)-
\pi'^{kl}(x)]\}} {(\det[g''^{kl}(x)])^{1/2}\,(\det[g'^{kl}(x)])^{1/2}}  
\bigg\}\bigg) \;\label{e3}\\  
&&\hskip0cm \equiv\<\pi'',g''|\pi',g'\>\;.\no\en
First, observe that the matrices $\gamma''$ and $\gamma'$ do {\it not} 
explicitly appear in (\ref{e3}); they have each been replaced by 
  \bn  g(x)\equiv e^{\gamma(x)/2}\,{\tilde g}(x)\,e^{\gamma(x)^T/2}
\equiv\{g_{ab}(x)\} \;,\en
where $\<\eta|\hg_{ab}(x)|\eta\>\equiv {\tilde g}_{ab}(x)\,
[={\tilde g}_{ba}(x)]$, a fixed reference metric that only serves 
to define the underlying topology of the space being quantized. 
Since only $g$ (and not $\gamma$) appears in the chosen functional 
form, we have renamed the overlap function $\<\pi'',g''|\pi',g'\>$ 
without loss of generality. This fact implies that the coherent states 
themselves are equally well denoted by $|\pi,g\>$. Second, note that 
the representation (\ref{e3}) is ultralocal, i.e., specifically of the form
  \bn  \exp\{-\tint b(x)\,d^3\!x\,L[\pi''(x),g''(x);\pi'(x),g'(x)]\,\}\;, \en
 and thus, {\it by design}, there are no correlations between or among 
spatially separated field values, a neutral position towards correlations 
before any constraints are introduced. On invariance  grounds, (\ref{e3}) 
necessarily involves a {\it scalar density} $b(x)$, $0< b(x)<\infty$, for 
all $x$; this arbitrary and nondynamical auxiliary function $b(x)$ is only 
{\it temporary} and it will disappear when the gravitational constraints 
are fully enforced, at which point proper field correlations will arise. 
In addition, note well that the coherent-state overlap functional is 
{\it invariant} under general spatial coordinate transformations. Third, 
and last, we emphasize that the expression $\<\pi'',g''|\pi',g'\>$ is a 
{\it continuous, positive-definite functional} and thus may be used as a 
{\it reproducing kernel} to define a {\it reproducing kernel Hilbert space} 
${\cal C}$ composed of continuous phase-space functionals $\psi(\pi,g)$ on 
which the initial, ultralocal representation of the affine field operators 
acts in a natural fashion. Some further explanation at this point may be 
helpful.
 
Although not commonly used, reproducing kernel Hilbert spaces are very 
natural and readily understood. By definition, the vectors $\{|\pi,g\>\}$ 
span the Hilbert space, and therefore elements of a dense set of vectors 
have the form 
   \bn  |\psi\>=\sum_{k=1}^K\alpha_k\,|\pi_{(k)},g_{(k)}\>\;, \en
for general sets $\{\alpha_k\}_{k=1}^K$ and $\{\pi_{(k)},g_{(k)}\}_{k=1}^K$, 
and some $K<\infty$. The inner product of two such vectors is clearly given 
by 
  \bn  \<\phi|\psi\>=\sum_{j,k=1}^{J,K}\s\beta^*_j\alpha_k\s\<\pi_{(j)},
g_{(j)}|\pi_{(k)},g_{(k)}\>\;. \label{e6} \en
To {\it represent} the abstract vectors themselves as functionals, we adopt 
the natural coherent-state representation, namely
  \bn \psi(\pi,g)\equiv \<\pi,g|\psi\>=\sum_{k=1}^K\alpha_k\,
\<\pi,g|\pi_{(k)},g_{(k)}\>\,.  \en
Thus, we have a {\it dense set of continuous functions}, $\{\psi(\pi,g)\}$, 
and {\it a definition of an inner product} between pairs of such functions, 
$(\phi,\psi)\equiv \<\phi|\psi\>$, as defined in (\ref{e6}). It only remains 
to complete the space to a (separable) Hilbert space $\cal C$ by adding the 
limit points of all Cauchy sequences in the norm $\|\psi\|\equiv(\psi,
\psi)^{1/2}$. Observe that these definitions imply that $(\<\cdot,
\cdot|\pi',g'\>,\psi)=\psi(\pi',g')$ and so the kernel $\<\pi,g|\pi',g'\>$ 
{\it reproduces} the original vector $\psi(\pi,g)$. Note well that {\it all 
properties} of the reproducing kernel Hilbert space $\cal C$ follow as 
direct consequences from just the reproducing kernel 
$\<\pi'',g''|\pi',g'\>$ itself; for details see, e.g., \cite{mesh}. 
$(\!\!($Of course, $f(y)=\int\s\delta(x-y)\s f(x)\,dx$ as well, but 
the difference is that the kernel $\delta(x-y)$ is neither continuous 
nor is an element of the Hilbert space of square integrable functions. 
Therefore, $L^2({\mathbb R})$ is {\it not} a reproducing kernel Hilbert 
space.$)\!\!)$

During the past several years, a functional integral formulation has been 
developed \cite{kla2} that, in effect, {\it within a single formula} 
captures the essence of {\it all four of the basic principles} described 
above. This ``Master Formula'' takes the form 
\bn  && \<\pi'',g''|\s\E\s|\pi',g'\>  \no\\
 &&\hskip1cm=\lim_{\nu\ra\infty}{\o{\cal N}}_\nu\s\int 
e^{-i\tint[g_{ab}{\dot\pi}^{ab}+N^aH_a+NH]\,d^3\!x\,dt}\no\\
  &&\hskip1.5cm\times\exp\{-(1/2\nu)\tint[b(x)^{-1}g_{ab}g_{cd}
{\dot\pi}^{bc}{\dot\pi}^{da}+b(x)g^{ab}g^{cd}{\dot g}_{bc}{\dot g}_{da}]\,
d^3\!x\,dt\}\no\\
  &&\hskip2cm\times[\Pi_{x,t}\,\Pi_{a\le b}\,d\pi^{ab}(x,t)\,
dg_{ab}(x,t)]\,\D R(N^a,N)\;. \label{e8} \en
Let us explain the meaning of (\ref{e8}). 

As an initial remark, let us 
artificially set $H_a=H=0$, and use the fact that $\int{\cal D}R(N^a,N)=1$. 
Then the result is that $\E=\one$, and the remaining functional integral 
yields the coherent state overlap $\<\pi'',g''|\pi',g'\>$ as given in 
(\ref{e3}). This is the state of affairs {\it before} the constraints are 
imposed, and remarks below regarding the properties of the functional 
integral on the right-hand side of (\ref{e8}) apply in this case as well. 
We next turn to the full content of (\ref{e8}).

The expression $\<\pi'',g''|\s\E\s|\pi',g'\>$  denotes the coherent state 
matrix elements of the projection operator $\E$ which projects onto a 
subspace of the original Hilbert space on which the quantum constraints 
are fulfilled in a
regularized fashion. Furthermore, the expression 
$\<\pi'',g''|\s\E\s|\pi',g'\>$ is another manifestly positive-definite 
functional that can be used as a reproducing kernel and thus used directly 
to generate the reproducing kernel physical Hilbert space on 
which the quantum constraints are fulfilled in a regularized manner. 
The right-hand side of equation (\ref{e8}) denotes a reasonably 
well-defined functional integral over fields $\pi^{ab}(x,t)$ and 
$g_{ab}(x,t)$, $0<t<T$, designed to calculate this important reproducing 
kernel for the regularized physical Hilbert space and which entails 
functional arguments defined by their smooth initial values 
$\pi^{ab}(x,0)=\pi'^{ab}(x)$ and $g_{ab}(x,0)=g'_{ab}(x)$ as well as 
their smooth final values $\pi^{ab}(x,T)=\pi''^{ab}(x)$ and 
$g_{ab}(x,T)=g''_{ab}(x)$, for all $x$ and all $a,b$. Up to a surface 
term, the phase factor in the functional integral represents the 
canonical action for general relativity, and specifically $N^a$ and 
$N$ denote Lagrange multiplier fields (classically interpreted as the 
shift and lapse), while $H_a$ and $H$ denote phase-space symbols (since 
$\hbar\ne0$) associated with the quantum diffeomorphism and Hamiltonian 
constraint field operators, respectively. The $\nu$-dependent factor in 
the integrand formally tends to unity in the limit $\nu\ra\infty$; but 
prior to that limit, the given expression {\it regularizes and essentially 
gives genuine meaning} to the heuristic, formal functional integral that 
would otherwise arise if such a factor were missing altogether \cite{kla2}. 
The functional form of the given regularizing factor ensures that the metric 
variables of integration {\it strictly fulfill} the positive-definite domain 
requirement. The given form and in particular the need for the nondynamical, 
nonvanishing, arbitrarily chosen scalar density $b(x)$, is very welcome  
since this form---{\it and quite possibly only this form}---leads to a
reproducing 
kernel Hilbert space for gravity having the needed infinite dimensionality; 
a seemingly natural alternative \cite{kla5} using $\sqrt{\det[g_{ab}(x)]}$ 
in place of
$b(x)$ fails to lead to a reproducing kernel Hilbert space with the required 
dimensionality \cite{wat2}. The choice of $b(x)$ determines a specific 
ultralocal representation for the basic affine field variables, but this 
unphysical and temporary representation {\it disappears entirely} after 
the gravitational constraints are fully enforced (as soluble examples 
explicitly demonstrate \cite{kla3}). The integration over the Lagrange 
multiplier fields ($N^a$ and $N$) involves a rather {\it specific 
measure} $R(N^a,N)$ (described in \cite{kla6}), which is normalized such 
that $\tint\D R(N^a,N)=1$. This measure is designed to enforce (a 
regularized version of) the
{\it quantum constraints$\s$}; it is manifestly {\bf not} chosen to enforce 
the classical constraints, even in a regularized form. The consequences of 
this choice are {\it profound} in that no (dynamical) gauge fixing is 
needed, no ghosts are required, no Dirac brackets are necessary, etc. In 
short, {\it no auxiliary structure of any kind is introduced}. (These 
facts are general properties of the projection operator method of dealing 
with constraints \cite{kla6,sch} and are not limited to gravity.) 

How one uses (\ref{e8}) to proceed further is detailed below, but the 
general idea, roughly speaking, is as follows. A major goal in the general 
analysis of (\ref{e8}) involves reducing the regularization imposed on the 
quantum constraints to its appropriate minimum value, and, in particular, 
for constraint operators that are partially second class, such as those of 
gravity, the proper minimum of the regularization parameter is 
{\it non\/}zero. Achieving this minimization involves {\it fundamental 
changes} of the representation of the basic kinematical operators, which, 
as models show \cite{kla3}, are so significant that any unphysical aspect 
of the original, ultralocal representation disappears completely. When the 
appropriate minimum regularization is achieved, then the quantum constraints 
are properly satisfied. The result is the reproducing kernel for the 
physical Hilbert space which then permits a variety of physical questions 
to be studied.

One may wonder why we have stressed the reproducing kernel definition of 
the inner product in the resultant Hilbert space and have not mentioned 
any of the standard integral expressions for forming inner products usually 
associated with coherent states. The reason for this is that {\it the 
functional representation of Hilbert space involved does {\bf not} possess 
a local integral expression to define the inner product of vectors despite 
the fact that it is a coherent state representation}. Such representations 
are said to involve {\it weak coherent states} \cite{klabost,kla97} and 
they are based on so-called nonsquare integrable representations of the 
associated group. Despite the lack of a usual local integral representation 
for the inner product, it is especially interesting that  there is, 
nevertheless, a nearly unchanged phase-space path integral representation. 
This fascinating and indeed somewhat surprising story arises for elementary 
systems as well and is described in \cite{kla97}, a paper that was 
motivated by the present study of quantum gravity.

\subsubsection*{Quantum constraints and their treatment}
The quantum gravitational constraints, $\H_a(x)$, $a=1,2,3$, and $\H(x)$, 
formally satisfy the commutation relations
 \bn &&[\H_a(x),\H_b(y)]=i\s\half\s[\delta_{,a}(x,y)\s\H_b(y)+
\delta_{,b}(x,y)\s\H_a(x)]\;,\no\\\
  &&\hskip.15cm[\H_a(x),\H(y)]=i\s\delta_{,a}(x,y)\s\H(y) \;,\\
  &&\hskip.31cm[\H(x),\H(y)]=i\s\half\s\delta_{,a}(x,y)\s[\s g^{ab}(x)
\s\H_b(x)+\H_b(x)\s g^{ab}(x) \no\\
&&\hskip3.6cm +g^{ab}(y)\s\H_b(y)+\H_b(y)\s g^{ab}(y)\s] \;. \no  \en
Following Dirac, we first suppose that $\H_a(x)\s|\psi\>_{phys}=0$ and 
$\H(x)\s|\psi\>_{phys}=0$ for all $x$ and $a$, where $|\psi\>_{phys}$ 
denotes a vector in the physical Hilbert space ${\frak H}_{phys}$. However, 
these conditions are {\it incompatible} since $[\H_b(x),g^{ab}(x)]\ne0$ and 
almost surely $g^{ab}(x)\s|\psi\>_{phys}\not\in{\frak H}_{phys}$, even 
when smeared. This means that the quantum gravity constraints are {\it 
partially second class}. While others may resist this conclusion, we 
accept it for what it is. One advantage of the projection operator method 
is that it treats all quantum constraints, e.g., first- and second-class 
constraints, on an {\it equal footing$\s$}; see \cite{kla6,sch}. In brief, 
if $\{\Phi_a\}$ denotes a set of self-adjoint quantum constraint operators, 
then
  \bn  \E=\E(\!\!(\Sigma\s\Phi_a^2\le\s\delta(\hbar)^2\s)\!\!) =
\int {\sf T}\s e^{-i\tint\lambda^a(t)\s\Phi_a\,dt}\,\D R(\lambda) 
\label{e10}\en
denotes a projection operator onto a regularized physical Hilbert space. 
Sometimes, just by reducing the regularization parameter $\delta(\hbar)^2$ 
to its appropriate size, the proper physical Hilbert space arises. Thus, 
e.g., if $\Sigma\s\Phi_a^2=J_1^2+J_2^2+J_3^2$, the Casimir operator of 
$su(2)$, then $0\le\delta(\hbar)^2<3\hbar^2/4$ works for this first class 
example. If $\Sigma\s\Phi_a^2=P^2+Q^2$, where $[Q,P]=i\hbar\one$, then 
$\hbar\le\delta(\hbar)^2<3\hbar$ covers this second class example. Other 
cases may be more involved but the principles are similar. The time-ordered 
integral representation for $\E$ given in (\ref{e10}) is useful in path 
integral representations and this expression is entirely analogous to the 
origin of $R(N^a,N)$ in (\ref{e8}). 
 
It is fundamentally important to make clear how Eq.~(\ref{e8}) was derived 
and how it is to be used \cite{kla2}. The left-hand side of (\ref{e8}) is 
an abstract operator construct in its entirety that came {\it first} and 
corresponds to one of the basic expressions one would like to calculate. 
The functional integral on the right-hand side of (\ref{e8}) came 
{\it second} and is a valid representation of the desired expression. 
However, the final goal is to turn that order around and to use the 
functional integral to {\it define and evaluate} (at least partially) 
the desired operator-defined expression on the left-hand side. In no way 
should it be thought that the functional integral was ``simply postulated'' 
as a ``guess as how one might represent the proper expression'', however 
suggestive that guess may be.

\subsection*{DIRECTIONS FOR FUTURE RESEARCH}
There are several directions for further research that are worthwhile, and 
let us elaborate on some of these. In a certain sense, they all take the 
Master 
Formula, Eq.~(\ref{e8}), as their starting point.

\subsubsection*{Reduced variable models}
To gain insight into (\ref{e8}) it is useful to study models with a finite 
number of degrees of freedom. This reduction, however, should be done so 
as to retain some of the second class nature of the quantum constraints 
characteristic of gravity. Most workers want to avoid quantum second class 
constraints at any cost assuming they must be ``wrong'' (i.e., unphysical). 
In fact, they arise as a direct consequence of the fundamentally different 
invariance groups of classical and quantum mechanics embodied in {\it 
fundamentally different} bracket relations, e.g., as given between generator 
elements (for $\hbar=1$) by
 \bn && \{e^{aq+bp},e^{cq+dp}\}=(ad-bc)\, e^{(a+c)q+(b+d)p}\;,  \\
    &&\hskip-.15cm[e^{aQ+bP},e^{cQ+dP}]=2i\s\sin[\half(ad-bc)]\, 
e^{(a+c)Q+(b+d)P}\;,   \en
where $a,b,c$, and $d$ represent free parameters.
So disturbing has the loss of full classical invariance been to some 
workers, it has prompted the program of ``Geometric quantization'' which 
changed the rules of operator representation so that the new quantum 
commutator bracket agreed in structure with the classical Poisson bracket. 
While such a modification is mathematically possible, it has essentially 
nothing to do with physics. 

We accept the implication that when a classical invariance is lost it is 
nevertheless replaced with a quantum invariance that reduces to the 
classical one as $\hbar\ra0$. Therefore, in the present philosophy, 
reduced variable models should be studied that retain the important 
feature of gravitational constraints that are first class classically 
and (partially) second class quantum mechanically. (Such features are 
avoided in traditional minisuperspace models.)

\subsubsection*{Metrical quantization} 
We have emphasized that (\ref{e8}) was first obtained by starting with the 
putative operator approach and deriving from it a valid functional integral 
representation. In so doing we have automatically been led to the 
all-important $\nu$-dependent continuous-time regularization factor that 
renders the functional integral representation so nearly well defined 
(see \cite{kla2} for details). In particular, observe that there is a 
{\it phase space metric} that appears in the regularization factor in 
(\ref{e8}). 

Alternative to how (\ref{e8}) was originally derived, one could accept the 
need for some such regularization and adopt the viewpoint that it is the 
{\it choice of a metric on gravitational phase space} that is the first and 
key ingredient in deciding how to initiate quantization by functional 
integration. This is the viewpoint of {\it Metrical quantization} 
\cite{kla9}. A preliminary study of alternative metrics has begun 
\cite{wat2}, and a more thorough examination is well warranted in order 
to determine whether the metric given in (\ref{e8}) is indeed optimal or 
perhaps other metrics are also worthy of study and would have the 
potential of leading to qualitatively different quantizations (as other 
examples demonstrate \cite{kla10}).  

\subsubsection*{Nonrenormalizability and symbols}
Viewed perturbatively, gravity is nonrenormalizable. However, the 
(nonperturbative) {\it hard-core picture of nonrenormalizability} 
\cite{kla11,book} holds that the nonlinearities in such theories are 
so strong that, from a functional integration point of view, a nonzero 
set of functional histories that were allowed in the support of the 
linear theory is now forbidden by the nonlinear interaction. This picture 
is qualitatively like that of a hard-core potential in nuclear physics 
which forbids particles from coming closer than a specified distance. In 
each case, expansion of the hard-core interaction in a pertubation series 
is surely inappropriate and alternative analyses are called for.

Various highly specialized field theory models exhibit analogous hard-core 
behavior and nevertheless possess suitable nonperturbative solutions 
\cite{book}. It is believed that gravity and also $\phi^4$ field theories 
in high enough spacetime dimensions can be understood in similar terms. A 
computer study to analyze the $\phi^4$ theory has begun, and there is hope 
to clarify that particular theory. Any progress in the scalar field case 
could strengthen the gravitational field case as well.

Evidence from soluble examples points to the appearance of a nonclassical 
(i.e., $\propto\hbar$) and nontraditional counterterm in the functional 
integral representing the irremovable effects of the hard core. These 
counterterms have an important role to play as part of the symbols 
representing the diffeomorphism and Hamiltonian constraints in the 
functional integral since for them $\hbar\ne0$ as well; indeed, in the 
chosen units $\hbar=1$. In brief, the form taken by the symbols $H_a$ 
and $H$ in (\ref{e8}) is intimately related to the proper understanding 
of how to handle the perturbative nonrenormalizability and the 
concomitant hard-core nature of the overall  theory. These are clearly 
difficult issues, but it is equally clear that they may be illuminated 
by studies of other nonrenormalizable models such as $\phi^4$ in 
sufficiently many dimensions. 

\subsubsection*{Classical limit}
Suppose one starts with a classical theory, quantizes it, and then takes 
the classical limit. It seems obvious that the classical theory obtained 
at the end should coincide with the classical theory one started with. 
However, there are counterexamples to this simple wisdom! For example, the 
$\phi^4$ theory in five spacetime dimensions has a {\it nontrivial} 
classical behavior. But, if one quantizes it by the lattice limit of a 
natural lattice formulation, the result is a free (or generalized free) 
quantum theory whose classical limit is also free and thus differs from 
the original theory \cite{fro}. This unsatisfactory behavior is yet another 
side of the nonrenormalizability puzzle. However, those nonrenormalizable 
models for which the quantum hard-core behavior has been accounted for do 
have satisfactory classical limits \cite{book}. The conjectured hard-core 
nature of $\phi^4$ models is under present investigation, and it is 
anticipated that a proper classical limit should arise. It is further 
conjectured that a favorable consequence of clarifying and including the 
hard-core behavior in gravity will ensure that the resultant quantum 
theory enjoys the correct classical limit. 

A few general remarks may be useful. It is a frequent misconception that 
passage to the classical limit requires that the parameter $\hbar\ra0$. 
To argue against this view, just note that the macroscopic world we know 
and describe so well by classical mechanics is the same real world in which 
$\hbar\ne0$. In point of fact, classical and quantum formalisms must 
{\it coexist}, and this coexistence is very well expressed with the help 
of coherent states. It is characteristic of coherent state formalisms that 
classical and quantum ``generators'', loosely speaking, are simply related 
to each other through the {\it Weak Correspondence Principle} \cite{corr}. 
In the case of the gravitational field, prior to the introduction of 
constraints, this connection takes the general form
  \bn  \<\pi,g|\s{\cal W}\s|\pi,g\>=W(\pi,g) \;.  \en
Here $\cal W$ denotes a quantum generator and $W(\pi,g)$ the corresponding 
classical generator (which is generally a ``symbol'' still since 
$\hbar\ne0$ ). The simplest example of this kind is given by $\<\pi,g|
\s\hg_{ab}(x)\s|\pi,g\>=g_{ab}(x)$. 

In soluble models where the appropriate classical limit has been obtained 
\cite{book}, coherent state methods were heavily used. It is expected that 
they will prove equally useful in the case of gravity.
 
\subsubsection*{Passage to the physical Hilbert space}

The reproducing kernel for the regularized physical Hilbert space given in 
(\ref{e8}) contains the seeds needed to define the reproducing kernel for 
the genuine physical Hilbert space. Rather than using (\ref{e8}) directly, 
however, we need to recognize that the Hamiltonian constraint, in 
particular, needs to be {\it regularized} since in its unregularized form 
it is incompatible with the original ultralocal representation of the basic 
kinematical operators. The regularization may be removed when the basic 
kinematical operators are in the proper representation, and the proper 
representation is determined by a proper choice of fiducial vector. In 
practice, we can say that to obtain a proper reproducing kernel for the 
physical Hilbert space one must {\it reduce the regularization parameter 
combined with any necessary rescaling of the reproducing kernel} as well 
as {\it recenter the reproducing kernel, which means to suitably change 
the fiducial vector}, all the while that one {\it removes the regularization 
introduced into the Hamiltonian constraint}.

One possible procedure to accomplish these goals would be as follows. 
Let us first introduce the notation
  \bn  U[(\pi,\gamma)]\equiv e^{i\tint \pi^{ab}\hg_{ab}\,d^3\!x}\,
e^{-i\tint\gamma^a_b\hp^b_a\,d^3\!x} \en
for elements of the affine group. Next let us define the parameter product 
$(\pi,\gamma)\cdot(\pi',\gamma')$ as the group  multiplication law so that
  \bn  U[(\pi,\gamma)\cdot(\pi',\gamma')]\equiv U[(\pi,\gamma)]\, 
U[(\pi',\gamma')] \;. \en
As a third step, let us form the states
 \bn &&|\pi,\gamma;\{a\}\>\equiv \sum_{k=1}^K\s a_k \,U[(\pi,\gamma)]\, 
U[(\pi_{(k)},\gamma_{(k)})]\,|\eta\> \no\\
   && \hskip1.9cm=\sum_{k=1}^K\s a_k \,U[(\pi,\gamma)\cdot(\pi_{(k)},
\gamma_{(k)})]\,|\eta\>\no\\
  &&\hskip1.9cm=\sum_{k=1}^K\s a_k \,|(\pi,\gamma)\cdot(\pi_{(k)},
\gamma_{(k)})\>\no\\ 
 && \hskip1.9cm\equiv\sum_{k=1}^K\s a_k \,|\pi_{((k))},\gamma_{((k))}\>\no\\
  &&\hskip1.9cm\equiv\sum_{k=1}^K\s a_k \,|\pi_{((k))},g_{((k))}\>  \;,  
\label{e18}\en
for suitable sets $\{a_k\}_{k=1}^K$ and $\{\pi_{((k))},g_{((k))}\}_{k=1}^K$, 
and some $K<\infty$. 
Observe, by this procedure, that we have been able to change the fiducial 
vector in the original reproducing kernel, since, according to the first 
line of (\ref{e18}), we see that
  \bn && |\pi,\gamma;\{a\}\>=  U[(\pi,\gamma)]\,\{\Sigma_{k=1}^K\s a_k\,
U[(\pi_{(k)},\gamma_{(k)})]\}\,|\eta\>\no\\
    && \hskip1.9cm\equiv U[(\pi,\gamma)]\,|\{a\}\> \label{f17}\;. \en

Let us denote by $\E_\Lambda$ the projection operator appearing in 
(\ref{e8}) where $\Lambda$ denotes the cutoff introduced  into the 
Hamiltonian constraint. Then, based on the earlier discussion, we can 
form a set of quotients determined by suitable linear sums of terms given 
by (\ref{e8}). Specifically, let us consider the set given by
\bn \bigg\{\frac{\<\pi'',\gamma'';\{a\}|\s\E_\Lambda\s|\pi',
\gamma';\{a\}\>}{\<0,{\tilde \gamma};\{a\}|\s\E_\Lambda\s|0,{\tilde 
\gamma};\{a\}\>}: {\rm for~all~nonzero}\; |\{a\}\>\; 
{\rm as~defined~in~(\ref{f17})}\;\bigg\}  \label{e19}\en
(It is assumed, of course, that no terms with a vanishing denominator 
are included.)
Among all elements of this set, one seeks to find those vectors 
$|\{a\}\>$ (i.e., those sets $\{a_k\}$ and $\{\pi_{((k))},g_{((k))}\}$) 
that maintain joint continuity in the arguments $(\pi'',\gamma'')$ and 
$(\pi',\gamma')$. As the cutoff is removed, i.e., as $\Lambda\ra\infty$, 
any given quotient will signal incompatibility with the current 
representation of the basic kinematical operators by losing joint 
continuity in the arguments $(\pi'',\gamma'')$ and $(\pi',\gamma')$. 
Ideally, to regain continuity one must reject the current operator 
representation and search for the proper representation, or provisionally 
at least, a more favorable representation. This search may be effected by 
examining various limits involving different fiducial vectors $|\{a\}\>$. 
In particular, exploring inequivalent operator representations involves 
changing Hilbert spaces, so to speak, and this entails taking suitable 
limits as $K\ra\infty$ that force the fiducial vector to leave its original 
Hilbert space. However, prior to taking all the required limits, 
indications of the proper direction to proceed should already be seen in 
the pattern of behavior within the original Hilbert space. This fact makes 
the study of the insipient discontinuity of various elements of (\ref{e19}) 
a vital clue to decide which direction to take when ``leaving the original 
Hilbert space''. 
 
This picturesque view can be made more precise and done so to an extent, 
hopefully, that some specific algorithm can be drafted which provides a 
practical way to find and study properties of changing the fiducial vector 
in an optimal fashion. Achieving the goal of finding the appropriate 
fiducial vector will lead to a new and proper representation of the basic 
kinematical operators that at the same time supersedes the original and 
provisional ultralocal representation. 

It is notable that the relevant variables of the resultant functional 
representation arise as a direct consequence of the very reduction procedure 
outlined above, an especially welcome consequence of using reproducing 
kernels. In particular, the physical reproducing kernel Hilbert space, 
generally speaking, has one or more ``spectator'' variables that are not 
necessary to span the Hilbert space, and even more importantly, they are 
not necessary to form the inner product in the associated reproducing 
kernel Hilbert space. It is among such variables that a suitable parameter 
to serve as ``time'' may be found for theories that originally were 
reparametrization invariant. Several soluble examples have confirmed 
this possibility; see, e.g., \cite{sch}. 

\subsection*{SUMMARY}
The approach to quantum gravity described in this article may seem unusual 
to the general reader, but it should be appreciated that its basic 
underlying principles are in fact quite conservative. A brief summary of 
the main points may be helpful.

Central to the present procedure is an insistence on the strict 
positive-definite character of the matrix of operators for the 
spatial metric. (A similar insistence does not seem to appear in either 
the canonical or string programs.) Compatibility with the metric spectrum 
requires using the affine commutation relations which are decidely 
{\it non\/}canonical and, instead, more like current algebras. Specific 
affine coherent states, projection operators to enforce constraints, and 
continuous-time regularized functional integral representations complete 
the formalism as presently constituted. Suitable limits to change initial 
operator representations have the potential of determining the proper 
physical Hilbert space for quantum gravity, and achieving that would enable 
many physical questions to be studied.

\subsection*{ACKNOWLEDGEMENTS}
Partial support for this work from NSF Grant PHY-0070650 is gratefully 
acknowledged.


\newpage

\begin{thebibliography}{99}
\bibitem{str}The many pages of: www.damtp.cam.ac.uk/user/gr/public/qg-ss.html.
\bibitem{ash}C. Rovelli, Living Reviews in Relativity, 
http://www.livingreviews.org/ Articles/Volume1/1998-1rovelli.
\bibitem{kla1}J.R. Klauder, ``Noncanonical Quantization of Gravity. I. 
Foundations of Affine Quantum Gravity'', J. Math. Phys. {\bf 40}, 5860 
(1999), gr-qc/9906013.  
\bibitem{kla2}J.R. Klauder, ``Noncanonical Quantization of Gravity. II.
Constraints and the Physical Hilbert Space'', J. Math. Phys. {\bf 42}, 4440 
(2001), gr-qc/0102041.
\bibitem{kla3}J.R. Klauder, ``Ultralocal Fields and their Relevance for
Reparametrization Invariant Quantum Field Theory'', J. Phys. A: Math. Gen. 
{\bf 34}, 3277 (2001), quant-ph/0012076.
\bibitem{mesh} N. Aronszajn, ``Th\'eorie G\'en\'erale de Noyaux 
Reproduisants - Premi\`ere Partie'', Proc. Cambridge Phil. Soc. 
{\bf 39}, 133 (1943); N. Aronszajn, ``Theory of Reproducing Kernels'',
Trans. Am. Math. Soc. {\bf 68}, 337 (1950); H. Meschkowski, 
{\it Hilbertsche Raume mit Kernfunktion}, (Springer Verlag, Berlin, 1962).
\bibitem{kla5} J.R. Klauder, ``Quantization = Geometry + Probability,'' in 
{\it Probabilistic Methods in Quantum Field Theory and Quantum Gravity}, 
Eds. P.H. Damgaard, H. H\"uffel, and A. Rosenblum, (North-Holland, 
Amsterdam, 1990), p. 73.
\bibitem{wat2} G. Watson and J.R. Klauder, ``Metric and Curvature in 
Gravitational Phase Space'', in preparation.
\bibitem{kla6} J.R. Klauder, ``Universal Procedure for Enforcing Quantum 
Constraints'', Nuclear Physics {\bf B 547},  397 (1999), hep-th/9901010.
\bibitem{sch} J.R. Klauder, ``Coherent State Quantization of Constraint 
Systems'', Ann. Phys. {\bf 254}, 419 (1997), quant-ph/9604033; J.R. 
Klauder, ``Quantization of Constrained Systems'', Lect. Notes Phys. 
{\bf 572}, 143 (2001), hep-th/0003297. 
\bibitem{klabost}J.R. Klauder and B.-S. Skagerstam, {\it Coherent States}, 
(World Scientific, Singapore, 1985). 
\bibitem{kla97}J.R. Klauder, ``Coherent State Path Integrals {\it without} 
Resolutions of Unity'', Found. Phys. {\bf 31}, 57 (2001), quant-ph/0008132.
\bibitem{kla9} J.R. Klauder, ``Metrical Quantization'', in 
{\it Quantum Future}, Eds. P. Blanchard and A. Jadczyk, (Springer-Verlag,
Berlin, 1999), p. 129, quant-ph/9804009.
\bibitem{kla10} J.R. Klauder, ``Quantization {\it Is} Geometry, After All", 
Ann. Phys. (NY) {\bf 188}, 120 (1988).
\bibitem{kla11}  J.R. Klauder, ``Field Structure Through Model Studies:  
Aspects of Nonrenormalizable Theories", Acta Physica Austriaca, Suppl. 
{\bf XI}, 341 (1973); J.R. Klauder, ``On the Meaning of a Nonrenormalizable 
Theory 
of Gravitation", Gen. Rel. Grav. {\bf 6}, 13 (1975); J.R. Klauder, 
``Continuous and Discontinuous Perturbations", Science {\bf 199}, 735 (1978). 
\bibitem{book}J.R. Klauder, {\it Beyond Conventional Quantization}, 
(Cambridge University Press, Cambridge, 2000). 
\bibitem{fro} R. Fern\'andez, J. Fr\"ohlich, and A.D. Sokal, {\it Random Walks
Critical Phenomena, and Triviality in Quantum Field Theory}, 
(Springer-Verlag, Berlin, 1992).
\bibitem{corr}J.R. Klauder, ``Weak Correspondence Principle", J. Math. 
Phys. {\bf 8}, 2392 (1967). 


\end{thebibliography}
\end{document}